\documentclass[12pt]{article}
\def \be{\begin {equation}}
\def \ee{\end {equation}}
\def \bea{\begin {eqnarray}}
\def \eea{\end {eqnarray}}
\def \intl{\int \limits}
\def \d{\partial}
\def \bphi{\mbox{\boldmath$\phi$}}

\begin{document}

\centerline{\Large  V.A. Sirota, A.S. Ilyin, K.P. Zybin, A.V. Gurevich. }
\bigskip

\begin{center}
\Large
The growth of supermassive black holes in the centers of galaxies:
absorption of stars and activity of galactic nuclei.
\end{center}

\begin{abstract}
The evolution of a supermassive black hole in the center of galaxy is considered.
We analyze  the  kinetic equation
describing relaxation processes associated with stellar encounters.
The initial distribution function of stars is assumed to be isothermal,
and it changes in accordance with consumption of stars by the black hole.
We calculate the consumption rate and obtain the law of growth of
supermassive black hole, i.e. the dependence of the black hole mass on time
and on the velocity dispersion of stars. The obtained values of mean black hole
mass and mean rate of stars consumption by a black hole are in reasonable
agreement with observational data.

\end{abstract}

{\bf 1. Introduction. }

The massive compact concentrations of mass in the centers of galaxies are
believed to be supermassive black holes with masses $10^6\div 10^9$ solar mass
($M_{\odot}$). The recent observations \cite{observ} provide significant
evidences for this hypothesis.

There are  two global approaches to the origin of these black holes:
they could either form together with the galaxy or even
earlier~\cite{Eroshenko}, or grow up during
galaxy evolution, capturing matter from the surrounding.
We throughout the paper consider the second approach.

Up to the present time there has been no consistent theory how the
growth of black holes in the galactic centers goes on.

Black hole can capture luminous (stars and gas) and cold dark matter from
the central parts of galactic core (bulge). Consumption rate of stars has
been investigated in \cite{Peebles, BahcWolf,Rees,LS,Dokuch}.
(The first statement of the problem of distribution and absorption of
particles in the Coulomb potential has been considered in \cite{Gurev}).
In \cite{Rees,LS,Dokuch} the diffusion in two-dimensional energy-momentum
space has first been discussed as the main process. However, all these authors
considered the dynamics of stars in the Coulomb potential of black hole,
whereas, as it has already been pointed out in \cite{LS}, in the case
of giant black holes  the consumption rate is determined by distances where
it is negligible and the stars move in their own self-consistent potential
field. On  the other hand, in all these papers stationary solution has been
considered; as we will show below, the solution of the real problem is
non-stationary.

The kinetic theory describing dynamics of non-baryon cold dark matter (CDM)
alone (without
baryon fraction and central black hole) has been developed in \cite{gz}.
It has been shown that the dark matter should have formed quasi-stationary
spherical structures with singular density distribution, bound by
gravitational forces. These structures are now observed as galactic halos.
The baryon matter should have fallen after recombination in the potential wells
formed by dark matter, beginning formation of galaxies
and forming seed black holes in the centers of the wells~\cite{gz90}.
This picture
explains well why the black holes are situated exactly in the dynamical centers
of galaxies~\cite{Cherep}: the potential is very smooth in the central part,
and  the black hole could not reach the center unless it would be formed just
in the center, in the potential well already existing.

The contemporary infall of CDM particles into the black hole caused by
gravitational interaction with stars and the corresponding growth of black
hole has been investigated in \cite{Anton}. It was shown that
this process leads to formation of  black holes
whose masses are consistent with the most part of supermassive black hole
masses measured nowadays.

In the present paper we analyze the mutual dynamics of cold dark matter and
stars in central parts of a galaxy. In accordance with previous papers, we
find that the most important process leading to the black hole growth is
(for both components) far encounters causing
 the relaxation in the energy-momentum space, or, to
be more precise, in the space of adiabatic invariants.
The influence of black hole appears both in gravitational potential (i.e.
the view of adiabatic invariants) and -- even stronger -- in boundary condition.

We derive the equations describing time-dependence of stars- and dark
matter- distribution functions, and calculate the coefficients in
one particular assumption of isothermal distribution of stars and the
potential dominated by stars, either (Section 2).

In Section 3 we discuss the boundary condition caused by presence of black
hole, and the influence of the black hole potential.

Further, we find an approximate solution of the equation for stars, ignoring
presence of dark matter and gas (Section 4). The result shows that black hole
mass should grow in time as $t^{1/2}$, and is proportional to $\sigma^{3/2}$,
where $\sigma$ is the rms velocity in the inner regions of bulge.

In discussion, we review briefly the results of the paper. We show that the
mean energy released due to stellar matter absorption is in reasonable agreement
with the observed AGN activity.

{\bf 2.  Two-component dynamics. }

We denote by $f_*({\bf r},{\bf v})$ and $f_d({\bf r},{\bf v})$ the
distribution functions of stars and dark matter, respectively.
This means that fraction of mass of sort $\alpha$ contained
inside the interval $d^3 r$, $d^3 v$ is
$f_{\alpha} d^3 r d^3 v$.

The mutual dynamics of dark matter particles and stars including 
diffusion in the momentum space is described by kinetic equation with
Landau scattering integral in the  right part \cite{LL10}:
$$
\frac{d f_{\alpha}}{d t} = St [f] \ ,
$$
or
\be         \label{both}
\frac{\d f_{\alpha}}{\d t} + {\bf v}\cdot {\bf \nabla} f_{\alpha} +
{\bf \nabla} \Psi \frac{\d f_{\alpha}}{\d {\bf v}}
= 2 \pi G^2 \Lambda   \frac{\d}{\d v_k}
\sum \limits_{\beta} \int d^3 v' w_{kp} \left(
M_{\beta} f_{\beta}  \frac{\d f_{\alpha}}{\d v_{p}} -
M_{\alpha} \frac{\d f_{\beta}}{\d v'_{p}} f_{\alpha}   \right)
\ee
Here greek indexes denote the sort of particles, $M_{\alpha}$ denotes
the mass of particles,
$f_{\alpha}({\bf r}, {\bf v})$ corresponds to scattered particles,
$f_{\beta}({\bf r},{\bf v'})$ corresponds to scatterers,
$$ w_{kp}=\left(u^2\delta_{kp}-u_k u_p\right)/u^3 \ , \qquad {\bf u}={\bf v'}-{\bf v} \ , $$
${\bf u}$ is the relative velocity, $\Lambda \sim 15$ is gravitational Coulomb logarithm.
The gravitational potential $\Psi({\bf r})$ is determined by Poisson law,
$$
{\bf \Delta} \Psi = 4 \pi G \sum \limits_{\alpha} \rho_{\alpha} ({\bf r})
 = 4 \pi G \sum \limits_{\alpha} \int f_{\alpha} ({\bf r}, {\bf v}) \, d^3 v  \ ,
$$
where $\rho_{\alpha} ({\bf r})$ is mass density of fraction $\alpha$.

Comparing the terms in (\ref{both})
one can see that scattering of both dark matter and stars
by stars is $M_*/ M_{dm}$ times more effective than scattering by
dark matter particles, and in the case of dark matter
 the second term is $M_{dm}/M_*$ times smaller than the
first. So we rewrite the equation (\ref{both}) in the form
\bea         \label{1}
\frac{d f_*}{d t} = 2 \pi G^2 M_*  \Lambda  \left( \frac{\d}{\d v_k}
W_{kp}
\frac{\d f_*}{\d v_{p}} - \frac{\d }{\d v_{k}} W_{k} f_* \right)    \\
\frac{d f_d}{d t} = 2 \pi G^2 M_*  \Lambda  \frac{\d}{\d v_k}   \label{2}
W_{kp} \frac{\d f_d}{\d v_{p}}
\eea
where
\be  \label{W}
W_{k} = \int \frac{\d f_*}{\d v'_{p}} w_{kp} d^3 v' \ , \\
W_{kp} = \int  f_*  w_{kp} d^3 v'   \ , \\
\ee

\vspace{0.5cm}
{\bf Rewriting equation in terms of adiabatic invariants }

If the potential is spherically symmetric, one can introduce
Hamiltonian variables  $({\bf I},\bphi)$,
where ${\bf I}=(I,m,m_z)$ are adiabatic invariants
($m$ is absolute value of angular momentum, $m_z$ is its projection on $z$ axis,
$I$ is radial action; all -- per unit of mass):
\bea     \label{defI}
I=\frac 1{2\pi} \oint v_r dr = \frac 1{\pi} \intl_{r_{-}(E,m)}^{r_{+}(E,m)}
\sqrt{2\left( E-\Psi(r)-\frac{m^2}{2r^2}\right)} dr \ ,         \\
E=\frac{v^2}2 + \Psi(r) \ , \quad {\bf m}={\bf r}\times {\bf v} \ , \quad
E-\Psi(r_{\pm})-\frac{m^2}{2 r_{\pm}^2} =0  \ ,             \nonumber
\eea
and $\bphi$ are corresponding angles.
One of the advantages of these variables is that the adiabatic invariant
$I$ of a particle (unlike the energy $E$) does not change
during slow changes of the potential.

In these variables, the left side of the kinetic equation does not
contain derivatives over action variables $I_k$~\cite{Anton,budker}:
$$
\frac{\partial f_{\alpha}({\bf I}, \bphi)}{\partial t}+\omega_k ({\bf I})
\frac{\partial f_{\alpha}}{\partial \phi_k}=St_{I,\phi}[f_{\alpha}]
$$
We now make use of the fact that the frequency of collisions is much
smaller than the frequency of orbital motion.
After averaging over the angle variables $\phi_k$
the second  term in the equation vanishes, and
the equations (\ref{1}),(\ref{2})
then take the form (see \cite{budker,Anton} for detailed derivation)
\be         \label{EquationI}
\frac{\d f_{\alpha}({\bf I})}{\d t} =  \frac{\d}{\d I_{k}} R_{kp}
\frac{\d f_{\alpha}}{\d I_p} -
\frac{M_{\alpha}}{M_*}\frac{\d }{\d I_k} (R_k f_{\alpha})  \ ,
\ee
where
\bea   
R_k = 2 \pi G^2 M_*  \Lambda
\frac 1{(2\pi)^3} \int d^3 \phi \frac{\d I_k}{\d v_p} W_p   \ ,
\nonumber \\  \label{Rkp}   \\ \nonumber
R_{kp} = 2 \pi G^2 M_*  \Lambda
\frac 1{(2\pi)^3} \int d^3 \phi
\frac{\d I_k}{\d v_m}\frac{\d I_p}{\d v_n} W_{mn}   \ .
\eea
We assume that in the vicinity of the center the distribution function
is spherically symmetric, i.e. does not depend on the direction of
angular momentum.  Then, there is a cylindrical symmetry in the space of
adiabatic invariants. Making use of
$$ R_{3}= \frac{m_z}{m} R_2  \ ; \quad
   R_{31}= \frac{m_z}{m} R_{21}  \ ; \quad
   R_{32}= \frac{m_z}{m} R_{22}  \ ,
$$
we can exclude $m_z$ from the equation~(\ref{EquationI}).
Eventually, we obtain the equations for distribution functions of
stars and of dark matter:
\be  \label{withoutmz}
\frac{\d f_*}{\d t} =
\frac{\d}{\d I} \left( R_{11}\frac{\d f_*}{\d I} +
 R_{12}\frac{\d f_*}{\d m} - R_1 f_* \right)   +
\frac 1m \frac{\d}{\d m} m
\left( R_{21} \frac{\d f_*}{\d I} + R_{22} \frac{\d f_*}{\d m} -R_2 f_* \right)
\ee
\be  \label{withoutmzdm}
\frac{\d f_d}{\d t} =
\frac{\d}{\d I} \left( R_{11}\frac{\d f_d}{\d I} +
 R_{12}\frac{\d f_d}{\d m} \right)   +
\frac 1m \frac{\d}{\d m} m
\left(  R_{21} \frac{\d f_d}{\d I} + R_{22} \frac{\d f_d}{\d m}  \right)
\ee
We see that evolution of distribution functions is described by axially
symmetric diffusion in three-dimensional space of adiabatic invariants.

We note that
equations (\ref{EquationI}), (\ref{withoutmz}) have a divergent form, i.e.
$$ \frac{df }{dt} =  \frac{\d}{\d I_k} S_k $$
The flux of matter through any surface in the momenta space
is given by the integral of ${\bf S}$ over the surface. For example,
the flux through the plane $I=0$ is
$\int dm \int dm_z S_I = \int 2 m S_I dm $. The boundary $I=0$ is
not physical, hence  flux of matter over any part of  it should be zero.
Thus, we obtain the boundary condition
\be        \label{firstbound}
\left. S_I \right| _{I=0} =
\left. \left( R_{11} \frac{\d f_{\alpha}}{\d I} +
R_{12} \frac{\d f_{\alpha}}{\d m} -R_1 f_{\alpha} \right)  \right|_{I=0} =0
\ee
In absence of a black hole, the other boundary condition would be
$$
\left. S_m \right| _{m=0} = 0 \ .
$$
The second boundary condition imposed by presence of black hole will
be considered in Section 3.

\vspace{0.5cm}
{\bf Isothermal distribution function }

One important example is the Maxwellian stars distribution function with
arbitrary time-independent potential:
$$
f_{M} = f_0 e^{-E/\sigma^2}
$$
This function corresponds to the equilibrium state,
since it satisfies equation (\ref{1}). This means
that both parts of equation (\ref{withoutmz}) are equal to zero. Moreover,
flux of matter over any surface in the momenta space
must be zero, hence all components of the flux ${\bf S}$ must vanish.
This puts two relations  on coefficients $R_k, R_{kp}$:
\be \label{sviazi}
R_{11} \frac{\d f_{M}}{\d I} + R_{12} \frac{\d f_{M}}{\d m} -R_1 f_{M} =0
\ , \quad
R_{21} \frac{\d f_{M}}{\d I} + R_{22} \frac{\d f_{M}}{\d m} -R_2 f_{M} =0
\ee

In the particular case of potential dominated by stars, there is the only
parameter $\sigma$ characterizing the distribution function:
\be
f_{is} = f_0 e^{-E/\sigma^2}  \ , \quad   \label{isoterm}
\Psi_{is} = 2 \sigma^2  \ln r \ , \quad
f_0=\left[ (2\pi)^{5/2} G \sigma \right] ^{-1} \ .
\ee
This function is called isothermal, the parameter $\sigma^2$ corresponds
to velocity dispersion.

We now calculate the coefficients $R_k, R_{kp}$ for isothermal distribution function,
$f_*=f_{is}$. All manipulations needed for the procedure
are described in Appendix. The result is
\bea
R_{22} \simeq 0.5 G M_* \Lambda \sigma \equiv R \ ;         \nonumber \\
R_{12} \simeq - 0.6 R (1-\mu \Phi_2(\mu) )  \ ;    \nonumber \\
R_{11} \simeq R (0.4 \Phi_1(\mu) - 0.8 \mu \Phi_2(\mu)+0.4 ) \ ;
\nonumber \\
R_1 \simeq - \frac{R}{I+0.6 m}
            \left( 0.7 \Phi_1(\mu) - 0.8 \mu \Phi_2(\mu) \right) \ ;
\label{chisla} \\
R_2 \simeq - 1.3 \frac{R}{I+0.6 m} \mu \Phi_2(\mu)   \ ; \nonumber \\
\mu \simeq 0.3 \frac{m}{I+0.6 m}   \nonumber
\eea
where approximate expressions for functions $\Phi$ are:
$$ \Phi_2(\mu) \simeq 1+ 0.35\mu^{-0.6} \ , $$
$$\Phi_1(\mu) \simeq 0.5 - 0.8 \ln \mu \cdot (\Phi_2 (\mu)-1)
\simeq 0.5 + 0.33 \mu^{-0.9}  \ .             $$

The isothermal distribution function can also be expressed in terms
of Hamiltonian variables (see Appendix):
\be  \label{fisoterm}
f_{is}(I,m) = f_0 e^{-E/\sigma^2}
 \simeq  \frac{ 0.3 f_0 \sigma^2 }{(I+\frac 2{\pi} m)^2}
\ee

\vspace{1.5cm}
{\bf 3.  The black hole, and  setting the problem.}

The scattering of dark matter particles by stars and their consumption
by a black hole has been discussed in \cite{Anton}. It has been assumed
in that paper that growth of black holes is fully determined by dark matter.
Here we discuss an opposite assumption of full absence of dark matter.
From now on, we discuss only dynamics of stars. Accordingly, we hereafter
omit the index $*$ in $f_*$.

Black hole can effect the dynamical evolution of surrounding matter in
two ways: consuming matter from the loss-cone and changing the
gravitational potential. We now consider the first effect
as principal one, and later on will briefly discuss the second.

The black hole with mass $M_{bh}$ disrupts (and after that consumes)
all stars that come nearer than the tidal radius
$ r_t = \left( \frac 6{\pi}  \frac{M_{bh}}{\rho_*} \right)^{1/3} $,
where $\rho_*$ is the density of the star.
We assume it to be approximately equal to solar density $\rho_{\odot}$.
In terms of variables $(I,m)$
this means that during one oscillation period all stars inside the loss-cone
\be        \label{mtidal}
m \le m_t = \sqrt{2 G M_{bh} r_t} \simeq 6\cdot 10^{18}
\left( \frac{M_{bh}}{M_{\odot}} \right)^{2/3}
\left( \frac{\rho_{\odot}}{\rho_*} \right)^{1/6}  \mbox{cm}^2/\mbox{s}
\ee
are disrupted. This boundary condition on $m$ has first been pointed out
in \cite{LS}.
On the other hand, if angular momentum is less than $m_g$:
\be   \label{mg}
m \le m_g = 4 G M_{bh} / c    \ ,
\ee
the star will anyway fall into the black hole \cite{LL2}, independently of
$m_t$. (If the black hole mass $M_{bh}$ is greater than
$\sim 3 \cdot 10^8 M_{\odot}$, the tidal radius becomes smaller than the
gravitational radius $r_g$, and stars become consumed without disruption.
For dark matter particles the loss cone is always defined by $m_g$.)
For brevity, we shall hereafter denote the boundary value by $m_t$,
assuming maximal of the two values $m_g$ and $m_t$:
$m_t = \mbox{max} \{ m_g, m_t \} $.

One can easily estimate the growth of the black hole caused by direct
capture of matter from the loss-cone:
taking, e.g., the isothermal function~(\ref{fisoterm}), we have
$ \Delta M_{bh} \sim 2 \pi \sigma m_t /G $. For black holes with masses
more than $10^3 M_{\odot}$ relative increase of mass $\Delta M_{bh}/M_{bh}$
is less than 10\% . Hence  the main growth is connected with diffusion of
matter into the loss-cone.

The condition that the particles vanish as they get into the area $m\le m_t$
may be expressed in terms of equation (\ref{withoutmz}) by boundary
condition
\be
\left. f \right|_{m=m_t} =0  \label{newbound}
\ee
It is the change of gradient of distribution function $\d f/\d m$ near
the boundary that causes shift of particles in the momenta space into the
loss-cone. This mechanism may lead to consuming much more mass than has
initially been inside the loss-cone.

Applying Gauss theorem to the equations (\ref{EquationI}),(\ref{withoutmz})
we see that the rate of black hole growth $d M_{bh}/dt$, i.e. loss
of whole mass of stars in the surrounding space, is equal to flux through the
boundary $m_t$:
\be                 \label{flux}
\dot{M}_{bh} = S =
\left. \int d^3 \phi
\intl_{0}^{\infty} dI \intl_{-m_t}^{+m_t} dm_z S_m \right|_{m=m_t}
  = \left. (2\pi)^3 2 m_t     \intl_{0}^{\infty} dI
  \left( R_{22} \frac{\d f}{\d m} +R_{12} \frac{\d f}{\d I}
  -R_2 f \right)    \right|_{m=m_t}
\ee
The last two terms are zero because of the boundary condition (\ref{newbound}).

So, to find the contemporary mass of black hole and distribution of stars
in its vicinity one should solve equation (\ref{withoutmz}) with
boundary conditions (\ref{firstbound}) and (\ref{newbound}).
The isothermal function (\ref{fisoterm}) is the most natural choice of
initial distribution because it is stationary
distribution with self-consistent potential and with no flows.
If there were no black hole, the isothermal
distribution remained constant. The boundary condition (\ref{newbound})
disturbs the equilibrium and makes the distribution change.

Furthermore, we take the same isothermal function to calculate the coefficients
$R_k, R_{kp}$ at any moment of time. We suppose that one may neglect the
changes in distribution of scatterers, because
(\ref{Rkp}) includes only integrated function
$f_*$, and changes in $f_*$ do not effect the coefficients significantly.
We will check the convenience of this approximation in Section 4.

Let us now estimate the influence of gravitational potential of
black hole. Taking values $R_{ik}$  of the isotherm (\ref{chisla})
for an estimate, one can see that to the lifetime of the Universe
$t\sim 3\cdot 10^{17}$s different members of (\ref{withoutmz}) may
produce significant changes of $f$ at scales $m \sim I \sim
\sqrt{R t} \sim 8\cdot 10^{25}$cm$^2$/s , this corresponds to \be
\label{ocenka}
r \sim 1 \mbox{pc} \cdot \left( \frac{\sigma}{200 \mbox{km/s}} \right)^{-1/2}
\ee
Dynamics of matter at these distances determines contemporary infall
into the black hole.
This is much smaller than  size of bulge ($\sim 1$kpc), so we should not
take boundaries of the distribution into account.
On the other hand, the black hole potential becomes significant at
distances
$$  r\le r_a \sim \frac{G M_{bh}}{\sigma^2} \sim 1 \mbox{pc}
\left( \frac{M_{bh}}{10^8 M_{\odot}} \right)
\left( \frac{\sigma}{200\mbox{km/s}} \right)^{-2}
$$
We see that for not very massive black holes,
\be    \label{ogrMpot}
M_{bh} \le 10^8 M_{\odot} \left( \frac{\sigma}{200km/s} \right)^{3/2} \ ,
\ee
the region of influence of the black hole potential does not
reach up to characteristic distances (\ref{ocenka}). In this case
only a small fraction of particles
enters the region, and, because of ellipticity of most orbits,
spends there rather small part of time. This is in full conformity with
conclusions by \cite{LS}.

The same arguments are applicable to the region
$ r< r_{coll} \sim R_{\odot} M_{bh} / M_{\odot} $ where orbital velocities
become comparable to the escape velocity from a typical star, and
inelastic scatterings are important \cite{Rees,LS}. One can easily
check that $r_{coll} \sim r_a$ for values $\sigma$ under discussion.

\vspace{0.5cm}
{\bf  4. Asymptotic solution.  }

The equation (\ref{withoutmz}) is difficult to solve. In this section
we obtain at least some integrated characteristics that make us
able to estimate the total consumption rate $S$ of stars and its evolution
in time.

Let us first discuss the vicinity of the black hole boundary,
$m_t \le m \ll \sqrt{R_{22} t}$
and $I > m$.
Comparing different terms in equation (\ref{withoutmz}) we find out
that, e.g., the term containing $R_{22}$ is of the order
of $R_{22}f/m^2 \gg \d f/\d t$, which means that the picture near the boundary
is quasi-stationary at contemporary time.
All the other terms except the two (with $R_{11},R_1$) that are
singular at $m\to 0$ (see (\ref{chisla})) are maximum of the order
of  $R_{22}f/Im$, i.e. much smaller.
To get rid of the two remaining terms, we integrate (\ref{withoutmz})
over $I$ and introduce $F(m)=\int f(I,m) dI$.
Thus, we obtain
\footnote{We neglect here small values of $I$, for which the terms with
$R_{12}$, $R_{2}$ may also be significant (although, no more than of the
same order as the term with $R_{22}$). However, it is easy to check that
in the vicinity of the boundary (and hence, everywhere) the adiabatic
invariant of each star could only grow with time (i.e., $S_I<0$), and
thus, contribution of small values $I$ in the flux $S$ and in the integrated
function $F$ is not large.   }
\be   \label{as1}
\frac 1m \frac{\d}{\d m} m R_{22} \frac{\d F}{\d m}   \simeq 0
\ee
Taking into account the boundary condition at $m=m_t$, we find
\be                                                     \label{asymptF}
F(m) = C(t) \ln \frac m{m_t}  \ , \quad m\ll \sqrt{R t}
\ee
The unperturbed isothermal function (\ref{fisoterm}) gives a solution
far from the black hole boundary,
$$                                               
F(m)=0.5 f_0 \sigma^2  \frac 1m  \ ,
              \quad m\ge \sqrt{R t}
$$
Substituting (\ref{asymptF}) into  (\ref{flux}), we see that
\be
S = (2\pi)^3 2 R_{22} C(t)         \label{asymflux}
\ee
The constraint on $C(t)$ is that the flux $S$ is equal to the whole loss
of mass:
$$
S = - \frac d{dt}     \int d^3 \phi \int dI \int dm \intl_{-m}^{+m} dm_z f =
   - (2\pi)^3 \frac d{dt}     \int 2 m F(m) dm
$$
Comparing this with (\ref{asymflux}) and calculating the integral,
assuming the two asymptotes are equal at $m=m_1$,
we obtain two equations:
\bea   \nonumber
C(t) m_1 \ln \frac {m_1}{m_t} = 0.5 f_0 \sigma^2  \ , \\
-\frac{\dot{C}}{C} m_1^2 \left( \ln \frac{m_1}{m_t} -\frac 12 \right) =2 R
\nonumber
\eea
If the boundary does not move, $m_t=const$, then
the approximate solution to the problem is
\bea
m_1(t) \simeq {2\sqrt{R t}}  \ln ^{-1/2} \left( \frac{2\sqrt{R t}}{m_t}\right)
                                           \label{solution} \\
C(t) \simeq \frac{f_0 \sigma^2 }
{4\sqrt{R t}} \ln ^{-1/2} \left( \frac{2\sqrt{R t}}{m_t} \right)
\eea
Now, knowing the dependence of flux $S = \dot{M}_{bh}$ on time, we calculate
the expression for mass of black hole.
$$
M_{bh} \simeq (2\pi)^3  f_0 \sigma^2
\sqrt{R t} \left/ \sqrt{\ln \frac{2\sqrt{R t}}{m_t}} \right.
$$
Substituting value of $f_0$, we find
\be
M_{bh} \simeq  \frac{2.6 \sigma \sqrt{R t}}
{G \sqrt{\ln \frac{2\sqrt{R t}}{m_t}}}               \label{Mbh}
\ee
The remarkable fact is that the result depends only weakly on the boundary
$m_t$. This is why the solution (\ref{solution})-(\ref{Mbh})
remains the same (up to the same accuracy) if we
account the change of $m_t$ with time ((\ref{mtidal}) or (\ref{mg}))
according to change of the black hole mass (\ref{Mbh}).

Taking $\Lambda=15$, $3\cdot 10^{17}$s -- the Universe lifetime,
and mass of star $M_*$ equal to solar mass, we obtain the mass of
black hole at arbitrary time:
\be M_{bh} \approx 1.2 \cdot 10^7
M_{\odot} \left( \frac{\sigma}{200 \mbox{km/s}} \right)^{3/2}
\sqrt{\frac{t}{3\cdot 10^{17} \mbox{s}}}            \label{finalM}
\ee
The contemporary consumption rate of stars
by black hole is (\ref{asymflux})
\be  \label{Sflux}
S \approx \frac{M_{bh}}{2t} \approx 6 \cdot 10^{-4} M_{\odot}/\mbox{year}
\cdot \left( \frac{\sigma}{200 \mbox{km/s}} \right)^{3/2} \ .
\ee
It decreases in time as $t^{-1/2}$, and corresponds to $3.6 \cdot
10^{43}$erg/s at contemporary time if $\sigma=200$km/s.
Occasionally, at $\sigma \simeq 200$km/s tidal and gravitational
limits of momentum $m_t$ and $m_g$ corresponding to the mass
(\ref{finalM}) are close to each other,    
$$
m_g \simeq     
2.7\cdot 10^{-3} \sqrt{R t} \frac{\sigma}{200 \mbox{km/s}}  \ ,
m_t \simeq     
4 \cdot 10^{-3} \sqrt{R t} \sqrt{\frac{\sigma}{200 \mbox{km/s}}}  \ .
$$
The corresponding  $m_1$ is
$$
 m_1 \simeq                0.8 \sqrt{R t} \ .
$$
We note that mass of black hole is no more than  half the mass
that was initially (when there was isothermal distribution) in the
area $m \le m_1$.   The same is correct for any time  (excluding the
very beginning).  This proves our initial
assumption that change of distribution function does not
affect significantly the coefficients $R_{kp}$.

\vspace{1cm}  {\bf Possible constraint.}

Let us now discuss briefly possible constraints on diffusion
model.
All equations beginning with~(\ref{EquationI}) are averaged over orbital period.
The boundary condition (\ref{newbound}) also assumes that all particles with
average angular momentum less than $m_t$ become captured into the black hole.
This means that during one period the change of trajectory is negligible.
But for very elongated and large trajectories change of angular momentum
of particle during one orbital period is comparable with the boundary angular
momentum $m_t$.
Hence, whether the star will or not be
captured by black hole during one period depends not on the mean
trajectory, but on its small fluctuations caused by individual
collisions.
\footnote{The effect was first pointed out in~\cite{BahcWolf,LS}. It was then
formulated in terms of critical energy, or critical radius in real space. }
In terms of adiabatic invariants this may be formulated as follows.

The diffusion approximation and equation (\ref{withoutmz}) are
applicable at any point of the momenta space. However, the
boundary condition (\ref{newbound}) may not be applied at very
large values $I$ (i.e., for very long orbits). Namely, mean
change of angular momentum $\Delta m$ during one period $T(I,m)$ is of
the order of $\Delta m \sim \sqrt{R T(I,m)}$, hence the constraint on
(\ref{newbound}) is
$$
\sqrt{R T(I,m_t)} \le \sqrt{R T(I_{crit},m_t)} = m_t
$$
On the other hand, we have seen in Section 3 that main changes in
the distribution function, and hence main contribution to the flux
occur at scales $I\le \sqrt{Rt}$ where $t$ is the age of the black
hole. This means that the consumption rate of stars is not
disturbed by influence of $I_{crit}$ if
$$ I\sim \sqrt{Rt} \ll I_{crit}$$.
At sufficiently large values $I\gg m_t$ the orbital period is
approximately equal to $T(I,m_t)\sim \pi I/\sigma^2$ (see
Appendix). Substituting $m_t$ from (\ref{mtidal}) we  obtain that
one anyway should not take the existence of $I_{crit}$ into account if
the black hole mass is greater than
\be \label{Mcrit}
M_{bh} \ge \left( \frac 1{50} \ \frac t{3\cdot 10^{17}\mbox{s}}
\frac{200\mbox{km/s}}{\sigma} \right)^{3/8} \cdot 10^7 M_{\odot}
\ee
Nowadays, this corresponds to minimal $M_{bh} \sim 2 \cdot 10^6
M_{\odot}$. Comparing the time dependence in (\ref{Mcrit}) with
that in (\ref{finalM}) we find that (if $\sigma$ has not changed
dramatically) the masses of the black holes that have grown as a
result of the described relaxation mechanism have always been greater than
the critical limit as far as galaxies exist. On the other hand,
for black holes with sufficiently smaller masses additional
investigations of influence produced by $I_{crit}$ are needed.

\vspace{1cm}
 {\bf 5. Discussion}

In the paper we derived the equations describing kinetic relaxation
of two-component (stars + dark matter) system in the vicinity of
supermassive black hole, in self-consistent gravitational potential.
The equations have been expressed in terms of Hamiltonian variables.

We have analyzed the growth of black hole caused
by consumption of stars, without taking non-baryon dark matter and
luminous gas into account.
We have seen that for black holes smaller than $10^6 M_{\odot}$ (\ref{Mcrit})
individual collisions near pericentre might play crucial role. For black
holes larger than $10^8 M_{\odot}$ (\ref{ogrMpot}) one should take
the potential of the black hole into account.
However, the most part of discovered black holes \cite{Cherep} have masses
in the range between $10^6 M_{\odot}$ and $10^8 M_{\odot}$ where
the main mechanism leading to consumption of stars is diffusion in the
momenta space, or in the space of adiabatic invariants.

Analogous mechanism has been discussed previously in a set of
papers \cite{Peebles,BahcWolf,Rees,LS,Dokuch}. But  all of them
have examined stationary distribution in the Coulomb potential of
the black hole. We have considered the evolution in
self-consistent potential and found that there is no stationary
state in presence of black hole. Also in previous papers only
diffusion along one axis (either $m$ or $E$) was taken into
account. The advantage of Hamiltonian variables $m,I$ is that in
this variables the diffusion along $m$ axis contains both shifts
along $m$- and $E$-axes in the $(E,m)$ plane and makes the main
contribution into the flux on the black hole. The diffusion along
$I$ axis plays less important role in the absorption process,
being merely responsible for change of form of stars trajectories.
Flux along $I$-axis increases smoothly as $m$ decreases, reaching
its maximum at the black hole boundary $m=m_t$, where
$S_I/S_m=-0.6$ (i.e., the particles in the $I-m$ plane intersect
the black hole boundary $m=m_t$ at the same angle independently on
$I$).

The solution (\ref{finalM}) obtained in Section 4 gives reasonable
estimate for contemporary characteristic black hole mass which lies in the middle
of observed range of masses. The estimate for flux of matter onto a black
hole (\ref{Sflux}) also corresponds to average observed activity of galactic
centers.

Equation (\ref{finalM}) also shows that $M_{bh}$ should change very slowly
in time, $M_{bh} \sim \sqrt{t} \sim (1+z)^{-3/4}$. Hence, the corresponding
difference between black hole masses in far and nearby galaxies is
too small and hidden by other effects.
On the other hand, the obtained  $M-\sigma$ relation differs from that
derived from observations
(from $\sim \sigma^{3.7}$ in \cite{Gebhardt}
to $\sim \sigma^{5.3}$ in \cite{MerrittFer})
However, this discrepancy is not critical, since black hole masses depend
not on one parameter $\sigma$ only, but also on the bulge mass \cite{Cherep},
mass of halo \cite{Baes, Anton}, amount of interstellar gas etc.
For example, the
contribution of dark matter could increase the resulting black hole mass
in those galaxies where amount of dark matter in the central parts is
significant.

In the previous paper \cite{Anton}  we have discussed the growth of black
hole caused by consumption of dark matter scattered by stars. The result was
that the characteristic black hole mass is
\begin{equation}\label{anton}
M_{bh (dark)}=8 \cdot 10^7 M_{\odot}
\left(\frac{M_H}{10^{12}M_\odot}\right)^{\frac{1}{2}}
\left(\frac{R_H}{100\mbox{kps}}\right)^{-\frac{9}{14}}
\left(\frac{\sigma}{200\mbox{km/s}}\right)^{\frac{4}{7}}
\left(\frac{t}{3 \cdot 10^{17}\mbox{s}}\right)^{\frac{4}{7}}  \ ,
\end{equation}
where $M_H$ and $R_H$  are the mass and the radius of the dark matter halo.

This result and solution~(\ref{finalM}) correspond to two limit cases describing
the black hole growth caused either by dark or by luminous matter consumption.
We see that contributions of stars and dark matter into the black hole
growth are comparable for galaxies where  densities of dark and luminous
matter in the bulge are of the same order. But dark matter plays in this
case a dominant role.
This, in particular, means that in galaxies where observations show very
small amount of dark matter the black hole mass should be in average smaller
than in others.
With account of real variations of all parameters ~(\ref{finalM})
and (\ref{anton}) cover the range
\be  \label{range}
5 \cdot 10^6 M_{\odot} \le M_{bh} \le 2 \cdot 10^8 M_{\odot}
\ee
which includes the most part of observed black holes.

The common point of view is that the consumption of stars by a black hole in
the galactic centre is the main source of energy providing activity of the
galactic nuclei \cite{Shapiro,MerFer}. To verify this hypothesis we
compared
the observed X-ray luminosity of the galactic centres with the central black
hole masses for 46 galaxies (Table 1). From the Table it follows that:
\begin{enumerate}
\item  The most part of the black holes have masses in the range (\ref{range}) restricted by
(\ref{finalM}) and (\ref{anton}).
\item  The observed intensity of X-ray emission is in a reasonable agreement
with the mean energy released as a result of stars consumption (\ref{Sflux}).
\item  The analysis of the data shows very strong variations in observed
luminosity of galactic nuclei. But no correlation with the
black hole masses was found. This indicates that the luminosity of one nucleus
should vary strongly in time. In other words, the duration $\tau_e$ of energy
emission following a star absorption is much less than the mean time interval
between two absorption events $T\sim M_{\odot}/ S \sim 2000$ years~(\ref{Sflux}).
From observed variations (3-4 orders of magnitude), assuming an
exponential fall of emission in time, we find an estimate $\tau_e \le T/10 \sim 200$years.
It follows that the most part of galaxies containing supermassive black holes are not active at
present (as, e.g., Milky Way). On the other hand, maximal intensity of
emission caused by star consumption should be much more than the average
value $S$ obtained in (\ref{Sflux}).
\item  In Fig.1, the histogram of average luminosity versus the black hole mass
is presented. The histogram is based on Table 1. It shows clearly that the
luminosity
breaks sharply at $M_{bh}>3\cdot 10^8 M_{\odot}$. This corresponds to the fact mentioned
in Section 3 that for such black holes the gravitational radius is greater than the tidal
radius, and stars sink into the black hole without disruption and energy emission. The observed cutoff
proves that it is the infall of stars -- not gas etc. -- that provides the main contribution
in the emission activity. On the contrary, in quasars with large redshifts
the main mechanism causing energy emission may be consumption of gas
\cite{quasars}.
\end{enumerate}

Thus, in this paper and in \cite{Anton} we present the theory of
growth of supermassive black holes in the galactic centers
caused by Coulomb scattering on stars and further consumption of
stars and dark matter. The theory is in reasonable agreement
with observations.
\vspace{0.7cm}

We thank Prof. A.M. Cherepashchuk and Prof. A.V. Zasov for
valuable discussions.
The research was supported by
RAS Program "Mathematical methods in nonlinear dynamics" and by the
President of RF grant "Support of Scientific schools" NSh-1603.2003.2.

\newpage
Table 1. \\
The X-ray luminosities and the central black hole masses of
galactic nuclei. \footnote{The black hole masses are adapted from
\cite{Cherep}, the X-ray luminosities -- from \cite{www}. Only
galaxies with measured X-ray luminosities are taken from the
sample of \cite{Cherep}. } 
\vspace{0.2cm}  \\
\begin{tabular}{lll}
Name               &    $L$, erg/s & $M_{bh}$, $M_{\odot}$ \\
3C 120  (Mrk 1506) &    $9.772\cdot 10^{43}$  &  $ 2.3 \cdot 10^7$        \\
                   &   $1.343\cdot 10^{44}$  &               \\
Ark 120 (Mrk 1095) &   $ 7.805\cdot 10^{43}$  & $  1.84 \cdot 10^8$       \\
Circinus           &   $ 2.47\cdot 10^{40}$   & $  1.3 \cdot 10^6$       \\
Fairall 9          &   $  6.688\cdot 10^{43}$ & $   8.0 \cdot 10^7$       \\
IC 1459            &   $ 8.398\cdot 10^{40}$  & $  3.7   \cdot 10^8$       \\
                   &   $ 1.051\cdot 10^{41}$  &               \\
IC 4329A           &   $  3.748\cdot 10^{43}$ & $   5 \cdot 10^6$       \\
                   &   $ 2.541\cdot 10^{43}$  &               \\
UGC 3973 (Mrk 79)  &   $ 2.594\cdot 10^{43}$  & $  5.2  \cdot 10^7$       \\
Mrk 110            &   $ 1.625\cdot 10^{44}$  & $  5.6 \cdot 10^6$       \\
Mrk 335            &   $ 2.04\cdot 10^{43}$   & $  6.3 \cdot 10^6$       \\
                   &   $ 4.916\cdot 10^{43}$  &               \\
Mrk 509            &   $ 1.188\cdot 10^{44}$  & $  5.78 \cdot 10^7$       \\
Mrk 590 (NGC 863)  &   $ 3.853\cdot 10^{43}$  & $  1.78 \cdot 10^7$       \\
                   &   $ 8.884\cdot 10^{43}$  &               \\
NGC 205 (M110)     & $<1.172\cdot 10^{38}$  & $ <9.3\cdot 10^{4}$  \\
NGC 598 (M33)      &   $ 1.46\cdot 10^{39}$    & $ <1.5\cdot 10^{3}$  \\
                   &   $ 1.902\cdot 10^{39}$   &               \\
NGC 1068 (M77)     &   $ 1.315\cdot 10^{41}$   & $  1.6 \cdot 10^7$        \\
                   &   $ 5.124\cdot 10^{41}$   &               \\
                   &   $ 8.81\cdot 10^{41}$    &               \\
NGC 3115           &   $ 1.773\cdot 10^{39}$   & $  9.1  \cdot 10^8$        \\
NGC 3227           &   $ 1.371\cdot 10^{42}$  &  $  3.9  \cdot 10^7$        \\
                   &   $ 7.127\cdot 10^{41}$   &            \\
NGC 3516           &   $ 5.578\cdot 10^{41}$   &  $ 2.3  \cdot 10^7$       \\
                   &   $ 1.084\cdot 10^{43}$  &               \\
NGC 3608           &   $ 1.174\cdot 10^{40}$  &  $ 1.1   \cdot 10^8$       \\
NGC 3783           &   $ 8.515\cdot 10^{42}$  &  $ 9.4  \cdot 10^6$      \\
                   &   $ 7.384\cdot 10^{42}$  &               \\
NGC 3998           &   $ 1.425\cdot 10^{41}$   &  $ 5.6 \cdot 10^8$         \\
NGC 4051           &   $ 7.609\cdot 10^{41}$   &  $ 1.3  \cdot 10^6$      \\
                   &   $ 7.624\cdot 10^{41}$   &               \\
NGC 4151           &   $ 9.025\cdot 10^{42}$  &  $ 1.53 \cdot 10^7$       \\
                   &   $ 5.578\cdot 10^{41}$   &               \\
                   &   $ 4.784\cdot 10^{41}$   &                 
\end{tabular}
\\
\begin{tabular}{lll}
NGC 4203           &   $ 7.471\cdot 10^{40}$  & $<1.2 \cdot 10^7$        \\
                   &   $ 6.424\cdot 10^{40}$  &                          \\
NGC 4258 (M106)    &   $ 2.914\cdot 10^{40}$  &  $4.1 \cdot 10^7$        \\
NGC 4261 (3C 270)  &   $ 1.575\cdot 10^{41}$   & $  5.2 \cdot 10^8$         \\
                   &   $ 1.434\cdot 10^{41}$   &               \\
NGC 4291           &   $ 6.128\cdot 10^{40}$  &  $ 1.5  \cdot 10^8$        \\
                   &   $ 6.769\cdot 10^{40}$  &               \\
                   &   $ 4.28\cdot 10^{40}$   &               \\
NGC 4342           &   $ 2.077\cdot 10^{39}$   & $  3.4  \cdot 10^8$       \\
NGC 4374 (M84)     &   $ 6.317\cdot 10^{40}$  &  $ 1.6  \cdot 10^9$       \\
                   &   $ 6.657\cdot 10^{40}$  &               \\
                   &   $ 6.138\cdot 10^{40}$  &               \\
NGC 4459           &   $ 4.169\cdot 10^{39}$   & $  6.5 \cdot 10^7$        \\
NGC 4473           &   $ 1.109\cdot 10^{40}$  &               \\
                   &   $ 5.089\cdot 10^{39}$   & $ 1.0   \cdot 10^8$        \\
NGC 4486 (M87)     &   $ 3.257\cdot 10^{42}$  & $ 3.4   \cdot 10^9$       \\
NGC 4593           &   $ 6.535\cdot 10^{42}$  & $  8.1  \cdot 10^6$      \\
NGC 4594 (M104)    &   $ 3.361\cdot 10^{40}$  & $  1.1  \cdot 10^9$       \\
NGC 4649           &   $ 1.002\cdot 10^{41}$   & $  2.0 \cdot 10^9$        \\
                   &   $ 1.587\cdot 10^{41}$   &               \\
NGC 4697           &   $ 9.141\cdot 10^{39}$   & $  1.2    \cdot 10^8$       \\
NGC 4945           &   $ 5.447\cdot 10^{39}$   & $ 1.1 \cdot 10^6$        \\
NGC 5548           &   $ 2.182\cdot 10^{43}$  &  $ 1.23 \cdot 10^8$       \\
                   &   $ 2.154\cdot 10^{43}$  &               \\
                   &   $ 2.778\cdot 10^{43}$  &               \\
NGC 6251           &   $ 1.71\cdot 10^{42}$   &  $ 5.4    \cdot 10^8$       \\
NGC 7469           &   $ 1.699\cdot 10^{43}$  &  $ 6.5 \cdot 10^6$       \\
                   &   $ 3.34\cdot 10^{43}$   &               \\
                   &   $ 2.071\cdot 10^{43}$  &               \\
PG 0026+129        &   $ 2.798\cdot 10^{44}$  &  $4.5 \cdot 10^7$         \\
PG 0052+251        &   $ 4.766\cdot 10^{44}$  &  $2.2  \cdot 10^8$         \\
PG 1211+143          & $ 1.098\cdot 10^{44}$  &  $ 4.05  \cdot 10^7$        \\
PG 1411+442          & $ 5.049\cdot 10^{42}$  &  $  8.0  \cdot 10^7$        \\
PG 1426+015(Mrk 1383)& $ 8.507\cdot 10^{43}$  &  $  4.7  \cdot 10^8$         \\
PG 1613+658(Mrk 876) & $ 1.652\cdot 10^{44}$  &  $  2.41  \cdot 10^8$       \\
PG 1617+175(Mrk 877) & $ 6.188\cdot 10^{43}$  &  $  2.73  \cdot 10^8$
\end{tabular}

\newpage

{\bf Appendix. \\
Calculating the  coefficients $R_k$,$R_{kp}$ in (\ref{Rkp})
and the adiabatic invariant in the case of isothermal stellar distribution
function.  }

A number of simplifications may be made for any isotropic distribution
function. We first  discuss the coefficients  $W_{kp}, W_k$
introduced in (\ref{W}).
If the distribution function $f_*$ is isotropic at any point of coordinate
space, i.e. does not depend on momentum,  but only on energy,
then tensor $W_{kp}$ depends on only one vector ${\bf v}$, and hence
takes the form
\begin{equation} \label{Wab}
W_{kp}=A(E,r)\delta_{kp}-B(E,r)\frac{v_k v_p}{v^2},
\end{equation}
where, calculating convolution $W_{kk}$ and product $W_{kp}v_p$, we find
\begin{eqnarray}
A=\frac{8\pi}{3} \intl_{\Psi(r)}^{\infty}dE'f(E')
\left\{
\begin{array}{rcl}
\frac{3}{2}\frac{v'}{v}\left(1-\frac{v'^2}{3v^2}\right)&,&E'<E\\
1&,&E'>E       \end{array}
\right.
\nonumber
\\
A-B=\frac{8\pi}{3} \intl_{\Psi(r)}^{\infty}dE'f(E')
\left\{
\begin{array}{rcl}
\frac{v'^3}{v^3}&,&E'<E\\  1&,&E'>E
\end{array}  \right.   \nonumber
\\
v=\sqrt{2(E-\Psi(r))} \ ,\quad v'=\sqrt{2(E'-\Psi(r))}  \nonumber
\end{eqnarray}
and similarly
\begin{eqnarray} \label{Walpha}
& W_{k} = \frac{v_{k}}v  D(E,r)  \ ,  \\
& D(E,r)= \frac{8\pi}3 \intl_{\Psi(r)}^{\infty}  dE' \frac{\d f}{\d E'} v
\left\{ \begin{array}{ll}
\left( v'/v \right)^3 \ , & E'<E \\ 1 \ , & E'>E  \end{array} \right. \nonumber
\end{eqnarray}

To calculate the coefficients $R_k$,$R_{kp}$ in (\ref{Rkp}) it is useful to
introduce new variables ${\bf \xi}=\{\xi_1,\xi_2\}=\{E,m\}$. Then, since
$I_i=I_i(E,m)$, we have from (\ref{Rkp})
$$  
R_{ij}=\frac{\d I_i}{\d \xi_a} \frac{\d I_j}{\d \xi_b} R'_{ab} \ , \quad
R_{i}=\frac{\d I_i}{\d \xi_a} R'_{a} \ ,
$$  
For brevity, we denote
$$ \Lambda_0 =2 \pi G^2 M_* \Lambda \ , \quad
<..>=\frac{1}{(2\pi)^3} \int d^3 \phi $$
Then, making use of (\ref{Wab}) and (\ref{Walpha}), we obtain the
expressions for $R'_a, R'_{ab}$:
\bea                    \nonumber
R'_{1}=\Lambda_0 < v D > \ ; \quad
R'_{2}=\Lambda_0 m < D/v > \ ; \\   \label{Rem} \\ \nonumber
R'_{11}=\Lambda_0 <v^2 (A-B)> \ ; \quad
R'_{12}=\Lambda_0 m <A-B>  \ ; \quad
R'_{22}=\Lambda_0 < A r^2 - B m^2/v^2 >
\eea
Note that because of spherical symmetry of all coefficients, only one
integral remains in the averaging:
$$
<..>= \frac 2{T(E,m)} \int \frac{dr}{v_r}
\ , \quad T(E,m)=\oint \frac{dr}{v_r} \ ,
\quad v_r=\sqrt{2\left( E-\Psi(r)-\frac{m^2}{2r^2}\right) }
$$

All the previous results are applicable for any isotropic function.
Now we substitute the isothermal distribution function (\ref{isoterm}).
For calculations it is useful to introduce dimensionless variables:
$$ \mu = \frac{m}{2 \sigma e^{E/2 \sigma^2}}   \ , \quad
x=r e^{-E/2 \sigma^2}  $$
The condition
$ v_r^2 = 2(E-\Psi - m^2/2 r^2) > 0  $ leads to
$$  0 \le x \le 1 \ , \quad 0 \le \mu \le 1/\sqrt{2e} \ .$$
Then, for example, period of oscillations is
$$
T(E,m) = \sqrt{2} \int
\frac{dr \theta(E-\Psi-m^2/2r^2)}{\sqrt{E-\Psi-m^2/2r^2 }} =
\frac 1{\sigma}  e^{E/2\sigma^2} \tilde{T} (\mu) \ , \quad
\tilde{T}(\mu) = \intl_0^1
\frac{dx \theta(-\ln x - \mu^2/x^2)}{\sqrt{-\ln x - \mu^2/x^2  }}
$$
The remarkable fact is that $\tilde{T} (\mu)$ is almost constant!
It grows very slowly from 1.77 at $\mu=0$ to 1.9 at $\mu \simeq 0.4$
- maximum possible $\mu$.

The averaging procedure now can be rewritten as
$$
<...>= \frac 2{T(E,m)} \intl_{r_-}^{r_+} \frac{dr}{v_r} (...)=
\frac 1{\tilde{T}(\mu)}  \intl_0^1
\frac{dx \theta(-\ln x - \mu^2/x^2)}{\sqrt{-\ln x - \mu^2/x^2  }}
(...)
$$
We also can calculate $A,B$ and $D$ for the isothermal distribution function:
$$
A = \frac{8\pi}3 \frac{f_0}{r^2} \sigma^2
\left( -2 \ln x (2H+x^2) -H+x^2 \right)  \ , \quad
A-B = \frac{8\pi}3 \frac{f_0}{r^2} \sigma^2
\left( 2 H+x^2 \right)  \ , $$
$$
D/v = - \frac{8\pi}3 \frac{f_0}{r^2} \left( 2 H+x^2 \right)  \ , \quad
v^2 = -4 \sigma^2 \ln x  \ , \quad
H(x)= \intl_{x}^1 y dy \sqrt{\frac{\ln y}{\ln x}}^3
$$
Note that
$$ A-B = -\sigma^2 D/v \  $$   which leads to (\ref{sviazi}).

From (\ref{Rem}) we now find
$$ R'_{11} =-\sigma^2 R'_1  \ , \quad  R'_{12} = -\sigma^2 R'_2 $$
$$ R'_1(E,m) = - 4\sigma^2 e^{-E/\sigma^2}
                \Lambda_0 \frac{8\pi}3 f_0 \Phi_1(\mu) $$
$$ R'_2(E,m) =  - m e^{-E/\sigma^2} \Lambda_0 \frac{8\pi}3 f_0 \Phi_2(\mu) $$
$$R'_{22}(E,m)= \sigma^2 \Lambda_0 \frac{8\pi}3 f_0 \Phi_{22}(\mu) $$
where
$$ \Phi_2 (\mu) = <2 H(x)/x^2 > +1 \ ,  \quad
 \Phi_1(\mu) = <-\ln x \cdot 2 H(x)/x^2> + <-\ln x>        $$
$$\Phi_{22} (\mu) = <-2\ln x (2H+x^2) -H+x^2> -\mu^2 <4H/x^2 + 1 +3H/(x^2\ln x)> $$
Up to at least $\mu \le 10^{-4}$ one can use approximate expressions:
$$ \Phi_2(\mu) \simeq 1+ 0.35\mu^{-0.6} \ , $$
$$\Phi_1(\mu) \simeq 0.5 - 0.8 \ln \mu \cdot (\Phi_2 (\mu)-1)
\simeq 0.5 - 0.3 \ln \mu \cdot \mu^{-0.6}               \ , $$
$$\Phi_{22}(\mu) \simeq 0.87$$

Finally, we calculate the adiabatic invariant $I$ (\ref{defI}):
$$
I(E,m) = \frac {2\sigma}{\pi} e^{E/2\sigma^2}
\intl_0^1 \sqrt{-\ln x - \mu^2/x^2} \theta(-\ln x - \mu^2/x^2) dx
$$
Calculating the integral we obtain
$$
I(E,m) \simeq \frac {2\sigma}{\pi} e^{E/2\sigma^2} (0.9-2\mu) =
\frac 2{\pi} (0.9 \sigma e^{E/2\sigma^2} - m)
$$
Hence,
$$    I'_m \simeq - \frac 2{\pi} \ , \quad
    I'_E \simeq \frac {I+\frac{2}{\pi}m}{2 \sigma^2} \ ;
$$
The distribution function can now be expressed in terms of adiabatic
invariants:
$$ 
f(I,m) = f_0 e^{-E/\sigma^2}
 \simeq  f_0 \sigma^2 \left( \frac{2 \cdot 0.9}{\pi} \right)^2
  \frac{1}{(I+\frac 2{\pi} m)^2}
$$
Another useful formula is the expression for the period $T(E,m)$
formulated in terms of $I,m$:
$$ T(I,m)=\frac{\tilde{T}(\mu)}{\sigma} e^{E/2 \sigma^2} \simeq
\frac 2{\sigma} \frac{\frac{\pi}{2} I + 2 m}{0.9 \sigma} \ . $$

\end{document}